\documentclass[%
 twocolumn,
 amsmath,amssymb,
 aps,
]{revtex4-2}

\usepackage{color}

\usepackage{graphicx}
\usepackage{dcolumn}
\usepackage{bm}

\begin{document}

\preprint{APS/123-QED}

\title{Lithium-cesium slow beam from a two-dimensional magneto-optical trap}

\author{Y.-D. Chen, W.-X. Li, M.-E. Chou, C.-H. Kuo, C.-S. Li, and S.~Tung}
\affiliation{%
 Department of Physics, National Tsing Hua University and Center for Quantum Technology, Hsinchu 30013, Taiwan
}%

\date{\today}

\begin{abstract}
We present the creation of a lithium-cesium slow beam using a two-dimensional magneto-optical trap. The two-species atomic beam is directed to load a three-dimensional magneto-optical trap in ultrahigh vacuum. We achieve a loading rate of 1.3$\times$10$^{7}$ lithium atoms/s with the lithium oven temperature at 370~$^{\circ}$C and 2.2$\times$10$^{7}$ cesium atoms/s with the Cs oven temperature at 20~$^{\circ}$C. The maximum numbers of lithium and cesium atoms in the trap are 1.0$\times$10$^{9}$ and 1.4$\times$10$^{8}$, respectively. Our results show that the simple and compact two-dimensional magneto-optical trap is suitable for producing an atomic beam of two species that have a large mass ratio and very different volatilities.

\end{abstract}

\maketitle

\section{\label{sec:level1}Introduction}

Mixtures of ultracold atoms offer unique possibilities to study broad and complex problems in few-body and many-body systems. The addition of a second species, isotope, or spin introduces new parameters that can have profound  influence on the properties of the systems. For example, the mass ratio of two species plays crucial roles in the discrete  scaling of Efimov states \cite{Ann.Phys.322.120.2007,PhysRevLett.112.250404,PhysRevLett.113.240402}, the dynamic properties of impurities and polarons \cite{PhysRevX.2.041020,PhysRevA.89.053617}, and the formation of quantum droplets \cite{PhysRevLett.115.155302,PhysRevA.98.053623,Science.359.301.2018,PhysRevLett.120.235301}. Furthermore, ultracold atoms of two species can be paired up to create polar molecules \cite{Ni231,PhysRevLett.114.205302,PhysRevLett.116.205303}. These molecules can be used as a platform for  quantum simulation \cite{NJP.11.055049.2009,Chem.Rev.112.5012}, precision measurements \cite{NJP.11.055049.2009}, and controlled chemistry \cite{NJP.11.055049.2009,Chem.Rev.112.4949}.

A lithium (Li) and cesium (Cs) combination offers interesting features. Lithium has two stable isotopes: $^{7}$Li (boson) and $^{6}$Li (fermion), while cesium has one: $^{133}$Cs (boson). $^{7}$Li-$^{133}$Cs and $^{6}$Li-$^{133}$Cs both have large mass ratios $\sim$20, and a Li-Cs system can be used to  study two-species degenerate gases, both Bose-Bose and Bose-Fermi. Furthermore, convenient Feshbach resonances exist in a $^{6}$Li-$^{133}$Cs mixture \cite{PhysRevA.87.010701,PhysRevA.87.010702} and in a $^{7}$Li-$^{133}$Cs mixture \cite{arXiv:2001.05329} as well, which allow tuning the interspecies interaction. Finally, a LiCs molecule has a large induced electric dipole moment of 5.5 D in its singlet ground state \cite{J.Chem.Phys.112.204302,PhysRevA.82.032503}, compared to 0.57 D for KRb \cite{Ni231}.

Having an efficient, robust atomic beam source is  essential for preparing mixtures of ultracold atoms in a magneto-optical trap (MOT). Zeeman slowers are commonly used to produce bright atomic beams. Inside a Zeeman slower, the magnetic field must satisfy an atomic mass-dependent profile to optimize the beam flux. Therefore, it is difficult to create a bright beam source simultaneously containing heavy and light atoms. Using a sequential loading scheme and integrating a second Zeeman slower are possible remedies to the problem. Both methods require extra laboratory work and become less practical when more species are included in the system.

Two-dimensional (2D) magneto-optical traps offer nice alternatives. Various 2D MOTs were first used to create slow beams of volatile species such as Cs \cite{Optics.communications.143.30,PhysRevA.60.R4241}, Rb \cite{PhysRevLett.77.3331,PhysRevA.58.3891}, and K \cite{PhysRevA.73.033415,J.Phys.B.amo.44.115307}, then later, of nonvolatile species such as Li \cite{PhysRevA.80.013409}, Na \cite{doi:10.1063/1.4808375}, and Sr \cite{PhysRevA.96.053415,PhysRevApplied.13.014013}. For producing single-species slow beams, a 2D MOT source's primary advantage is its smaller size. In the production of two- or multiple-species slow beams, a 2D MOT can simultaneously address each species, with relative ease.

In this paper, we report the creation of a lithium-cesium slow beam. We show that a two-species slow beam containing atoms of one volatile and one nonvolatile species can be efficiently produced using a single 2D MOT. Our setup is a modification of the 2D MOT design first reported in Tiecke $\it{et}$ $\it{al.}$ \cite{PhysRevA.80.013409} and Lamporesi $\it{et}$ $\it{al.}$ \cite{doi:10.1063/1.4808375}. Compared to Zeeman slowers, a two-species 2D MOT provides a simple, compact alternative to producing a simultaneous and overlapped slow beam regardless of the volatilities and the mass ratio of the two species. 

The paper is organized as follows. In Sec.~II we describe the experimental setup. The characterization and optimization of the two-species atomic beam are reported in Sec. III. In Sec. IV, we conclude with a brief discussion on the features of the beam source design.

\section{Experimental setup}
\subsection{Vacuum system}

\begin{figure}[h]
\centering
   \includegraphics[width=0.4\textwidth]{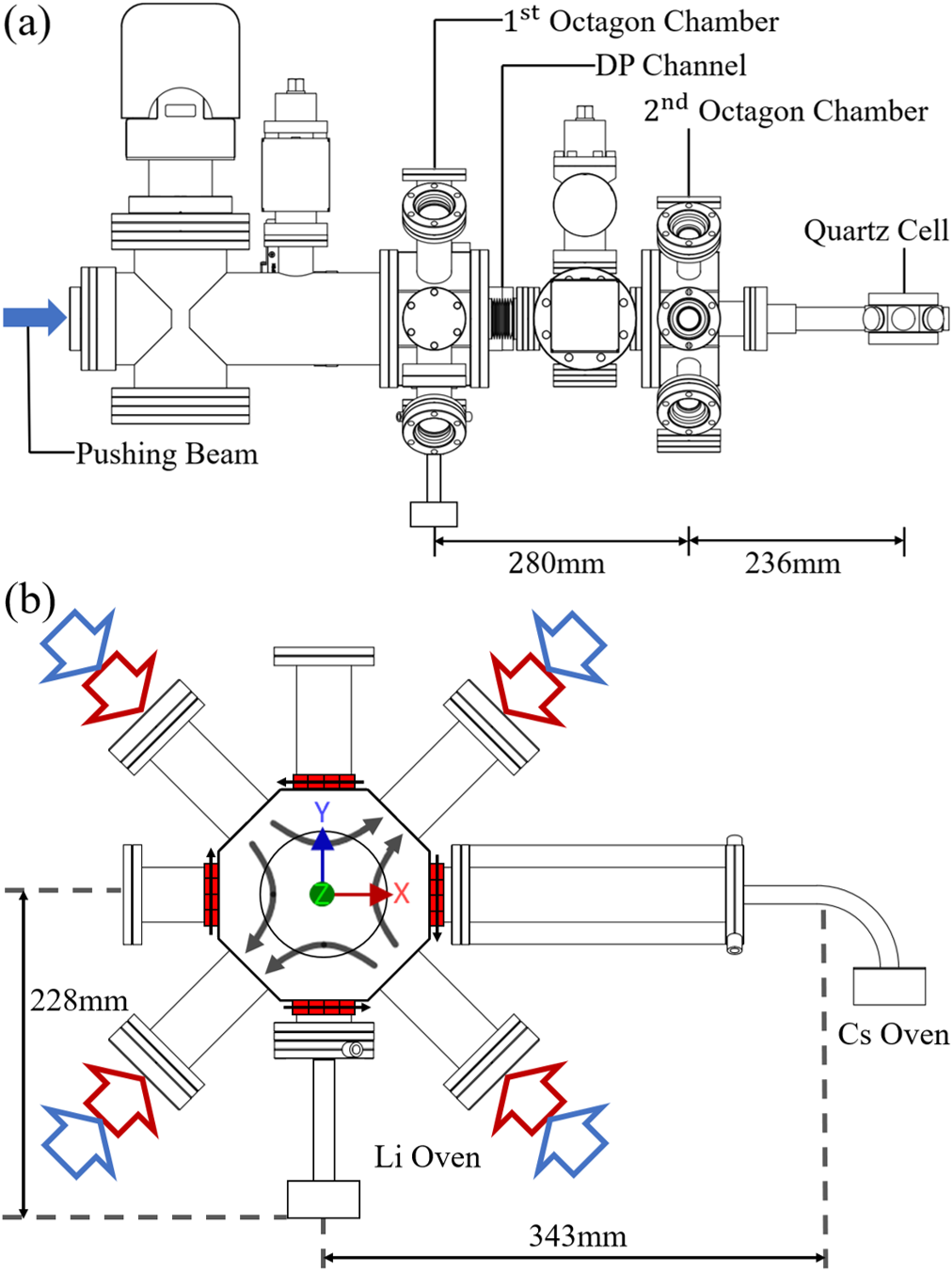}
   \caption{(Color Online) Schematic of the experimental setup: (a) the vacuum system consists of three major parts: two octagon chambers and a quartz cell. (b) Transverse view of the first octagon chamber. The Li and Cs ovens are attached to the bottom and the side of the chamber, respectively. The Li (red arrows) and Cs (blue arrows) laser beams cross at the center of the chamber, creating a cooling region. Eight stacks of permanent magnets (red blocks; only four stacks are visible in the schematic) are mounted at the positions ($\pm$83~mm, 0, $\pm$30~mm) and (0, $\pm$83~mm, $\pm$30~mm). They produce a 2D quadrupole magnetic field near the center of the chamber. The laser beams combined with the magnetic field create a two-species 2D MOT. The thin arrows on the magnets represent the directions of the magnetic dipoles, and the thick arrows indicate the directions of magnetic field. }
   \label{fig:vacuum system}
\end{figure}

A schematic representation of the experimental setup is shown in Fig.~\ref{fig:vacuum system}(a). The vacuum system includes three major components, two stainless steel octagon chambers and a quartz cell. A differential pumping (DP) channel, with a diameter of 2 mm and a length of 20 mm, is inserted to divide the system into high vacuum and ultrahigh vacuum regions. The high vacuum region includes the first octagon chamber, and the ultrahigh vacuum region contains the second octagon chamber and the quartz cell.
The first octagon chamber with Li and Cs ovens is the source chamber containing a two-species 2D MOT. In this work, the second octagon chamber is used for fluorescence detection. In the future, we plan to install a second 2D MOT in the chamber to refocus the atomic beam in order to increase the loading efficiency of the three-dimensional (3D) MOT located in the quartz cell.

\subsection{Ovens}
The Li oven has the shape of a cylinder, with a diameter of 50~mm and a height of 28~mm, and connects to a water-cooled CF~35 flange with a 87-mm-long tube which has an inner diameter of 11~mm. The distance from the bottom of the oven to the center of the first octagon chamber is 228~mm. The Cs oven has the same dimensions as the lithium oven, but it is mounted perpendicularly with respect to the Li oven. The oven is connected to the chamber by a 90$^{\circ}$ tube, with a length of 130~mm and an inner diameter of 11~mm. The distance from the 90$^{\circ}$ tube to the center of the 2D MOT is 343~mm [see Fig.~\ref{fig:vacuum system}(b)].

The Cs oven is loaded with 10~g of cesium, and the Li oven is loaded with 5~g of lithium, containing both $^{7}$Li and $^{6}$Li; $^{7}$Li and $^{6}$Li have natural abundances of $\sim$92.5$\%$ and $\sim$7.5$\%$, respectively. At room temperatures, the Cs vapor pressure is about 10$^{-6}$~mbar, while the Li vapor pressure is negligible in a conventional ultrahigh vacuum system; thus, when the system is kept in standby mode, we cool down the Cs oven to -15$^\circ$C, while maintaining the Li oven at 150 $^\circ$C. The corresponding vapor pressures are $\sim$10$^{-8}$~mbar for Cs and $\sim$10$^{-11}$~mbar for Li. To operate the system, we turn off the Cs oven cooler and heat the 90$^{\circ}$ tube to 100~$^\circ$C and the Cs oven to 20~$^\circ$C, while the Li oven is normally heated to 370$^\circ$C.
\subsection{Laser system}
\begin{figure}[h]
\centering
\includegraphics[width=0.45\textwidth]{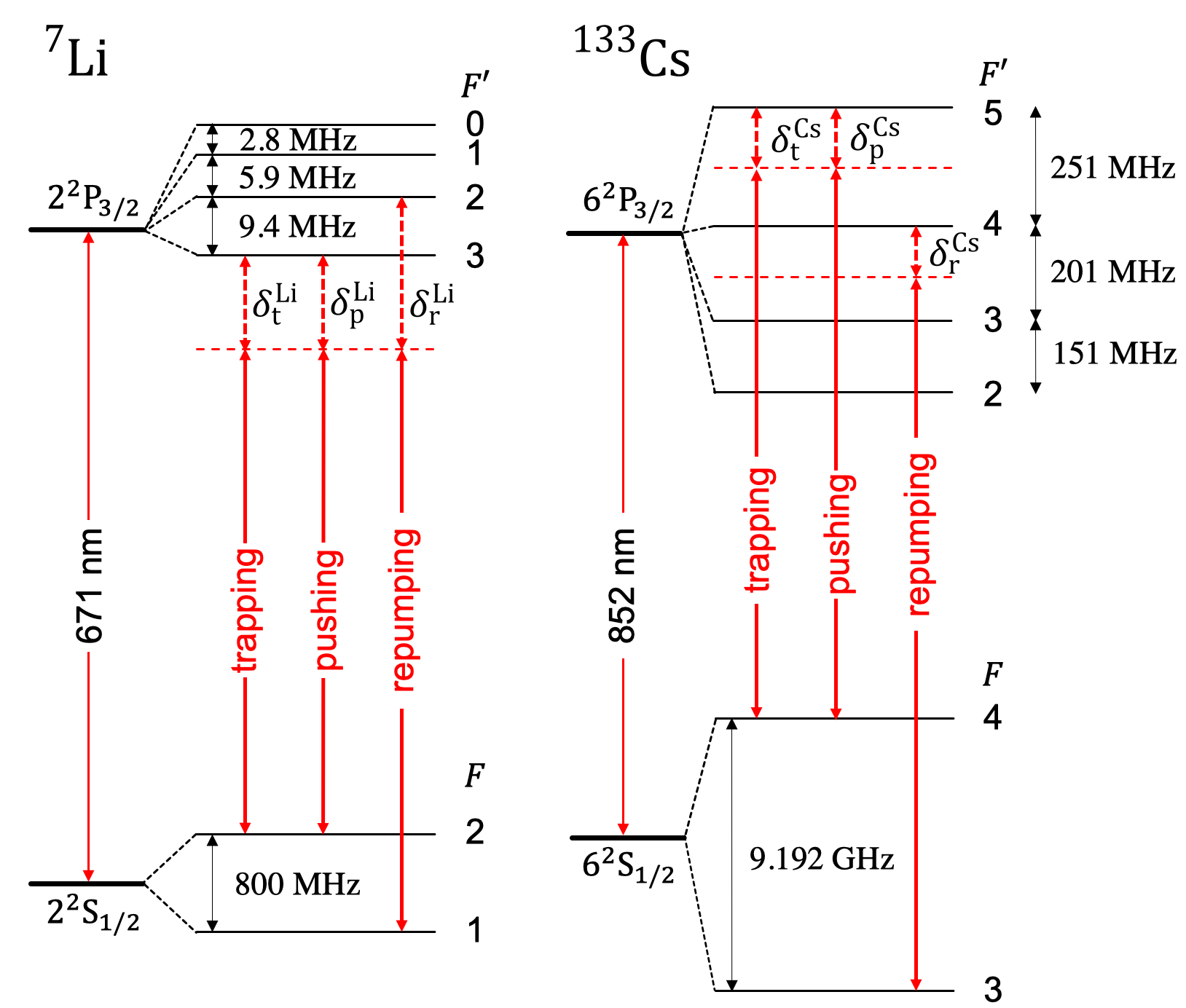}
\caption{(Color Online) Level diagrams of $^{7}$Li and $^{133}$Cs (not to scale) and laser frequencies used in the production of the Li-Cs slow beam. }

\label{fig:energy level} 
\end{figure}

Both Li and Cs possess strong, closed $S \rightarrow P$ transitions that are convenient for laser cooling and trapping. In this work, the cooling transitions are $\vert{\textrm{Cs}:F} = 4\rangle \rightarrow \vert{\textrm{Cs}:F'} = 5\rangle$  and $\vert{^{7}\textrm{Li}:F} = 2\rangle \rightarrow \vert {^{7}\textrm{Li}:F'} = 3\rangle$, and the repumping transitions are $\vert{\textrm{Cs}:F} = 3\rangle \rightarrow \vert{\textrm{Cs}:F'} = 4\rangle$ and $\vert{^{7}\textrm{Li}:F} = 1\rangle \rightarrow \vert{^{7}\textrm{Li}:F'} = 2\rangle$. Figure~\ref{fig:energy level} shows the energy levels and relevant laser frequencies for Li and Cs.

The Cs laser system consists of two external cavity diode lasers (Cs-$\alpha$ and Cs-$\beta$) and a tapered amplifier laser system (Cs-TA). The frequency of Cs-$\alpha$ is locked to the saturated absorption crossover of $\vert {\textrm{Cs}:F} = 4\rangle \rightarrow \vert{\textrm{Cs}:F'} = 3\rangle$ and $\vert{\textrm{Cs}:F} = 4\rangle \rightarrow \vert{\textrm{Cs}:F'} = 5\rangle$, while the Cs-$\beta$ is locked to the absorption resonance of $\vert{\textrm{Cs}:F} = 3\rangle \rightarrow \vert{\textrm{Cs}:F'} = 4\rangle$. An optical phase lock loop \cite{Appel_2009} is used to lock the frequency of Cs-TA 104~MHz above Cs-$\alpha$. Afterwards, acousto-optic modulators (AOMs) are employed to tune the Cs laser beams to the required frequencies shown in Fig.~\ref{fig:energy level}.

The Li laser system consists of two external cavity diode lasers (Li-$\alpha$ and Li-$\beta$) and two tapered amplifier laser systems (Li-TA1 and Li-TA2). Instead of locking to a Li frequency reference, we lock Li-$\alpha$ to the $a_{15}$ hyperfine component at transition R(78) 4-6 of $^{127}I_2$; the absolute frequency of the transition is measured to be 446,807,072.638(24)~MHz \cite{Huang:13}. Li-$\beta$, Li-TA1, and Li-TA2 are offset locked to Li-$\alpha$ and yield laser frequencies of 85.1~MHz below the resonant trapping transition, 251.2~MHz below the resonant trapping transition, and 232.7~MHz below the resonant repumping transition, respectively. After the Li lasers are locked, AOMs are used to tune the Li laser beams to the required frequencies shown in Fig.~\ref{fig:energy level}.

\subsection{Two-species 2D MOT}

A schematic drawing of the two-species 2D MOT is shown in Fig.~\ref{fig:vacuum system}(b). The 2D MOT is created by a 2D quadrupole magnetic field and two orthogonal pairs of overlapped, retroreflected Li and Cs laser beams. The diameters of the Li and Cs beams are 21 and 22~mm, respectively. Each laser beam contains trapping and repumping light required by laser cooling and trapping. Each Li beam has an average intensity of 16.3 mW/cm$^{2}$ with a trapping to repumping intensity ratio of 1.8, while each Cs beam has an average intensity of 19.8 mW/cm$^{2}$ with a trapping to repumping intensity ratio of 17. 

The 2D quadrupole magnetic field is produced by eight stacks of neodymium bar magnets. Taking the center of the chamber as the origin, the positions of the eight magnet stacks are ($\pm$83~mm, 0, $\pm$30~mm) and (0, $\pm$83~mm, $\pm$30~mm). A magnetic field gradient of 17, 31, and 40 G/cm is obtained with two, four, and six bar magnets in each stack, respectively. 

The 2D MOT cools hot atoms from the ovens. The longitudinal ($\it{z}$) velocity $v_z$ of atoms remains unchanged during the cooling process. Thus, atoms with a large $v_z$ do not stay in the laser fields long enough to be sufficiently cooled and collimated. These atoms will not be guided through the DP channel and loaded into the 3D MOT. Atoms with a small $v_z$ can be cooled effectively and form a beam propagating along the $\it{z}$ axis. However, these atoms  require a long traveling time to arrive at the 3D MOT, resulting in two effects that diminish the 3D MOT loading. First, since atoms have a finite transverse velocity, the transverse beam size can expand and become much larger than the trapping region of the 3D MOT; the effect is magnified for lighter species, such as lithium. Second, atoms can miss the 3D MOT because of gravity.

The velocity $v_z$ is a crucial parameter to optimize 3D MOT loading. Extra laser beams aligned with the $z$ axis, i.e., pushing beams, are used to accelerate the atoms towards the 3D MOT. We can control $v_z$ by tuning the detuning and the intensity of pushing beam.

\subsection{Two-species 3D MOT}
The two-species 3D MOT resides in the quartz cell, separated from the 2D MOT by a distance of $\sim$520~mm. The Li and Cs MOT beams are overlapped before being sent into the quartz cell. Each MOT beam contains two components of light:~trapping and repumping. The Li (Cs) trapping and repumping light in each beam have average intensities of 6.9(2.3) and 5.2(0.31)~mW/cm$^{2}$ with detunings of $-6$~$\Gamma_{\mathrm{Li}}$($-2.8$~$\Gamma_{\mathrm{Cs}}$) and $-3.5$~$\Gamma_{\mathrm{Li}}$(0~$\Gamma_{\mathrm{Cs}}$), respectively. Here, $\Gamma_{\mathrm{Li}}$($\Gamma_{\mathrm{Cs}}$) is the natural linewidth of the Li(Cs) $D_\mathrm{2}$ transition. The Li and Cs MOT beams have a similar diameter of 23~mm. In the 3D MOT, the quadrupole magnetic field is generated by a pair of electromagnetic coils. The measurements shown in the paper are taken with fixed field gradients:~${\partial B_y}/{\partial y}$=2(${\partial B_x}/{\partial x}$)=2(${\partial B_z}/{\partial z}$)=15~G/cm.

\section{Results and Discussion}
\subsection{Atomic beam characterization}

We characterize the properties of the two-species atomic beam using a time-of-flight (TOF) method \cite{PhysRevA.58.3891}. This is implemented by illuminating the atomic beam with resonant probe beams and then imaging the fluorescence on a photodiode after the 2D MOT is shut off abruptly. The Li and Cs probe beams are both elliptical and have a beam size of 10~mm $\times~$2 mm ($1/e^2$ diameter). The probe beams are aligned to the center of the second octagon chamber with the long axis perpendicular to the atomic beam and retroreflected to prevent the atoms from being pushed out of resonance. The average intensities of the Li and Cs probe beams are 27.7 and 21.3~mW/cm$^{2}$ with resonant trapping~($\delta_{t}^{\textrm{Li}}$=$\delta_{t}^{\textrm{Cs}}$=0) to repumping~($\delta_{r}^{\textrm{Li}}$=$\delta_{r}^{\textrm{Cs}}$=0) light ratios of 0.9 and 90, respectively.

\begin{figure}[h]
    \centering
    \includegraphics[width=0.45\textwidth]{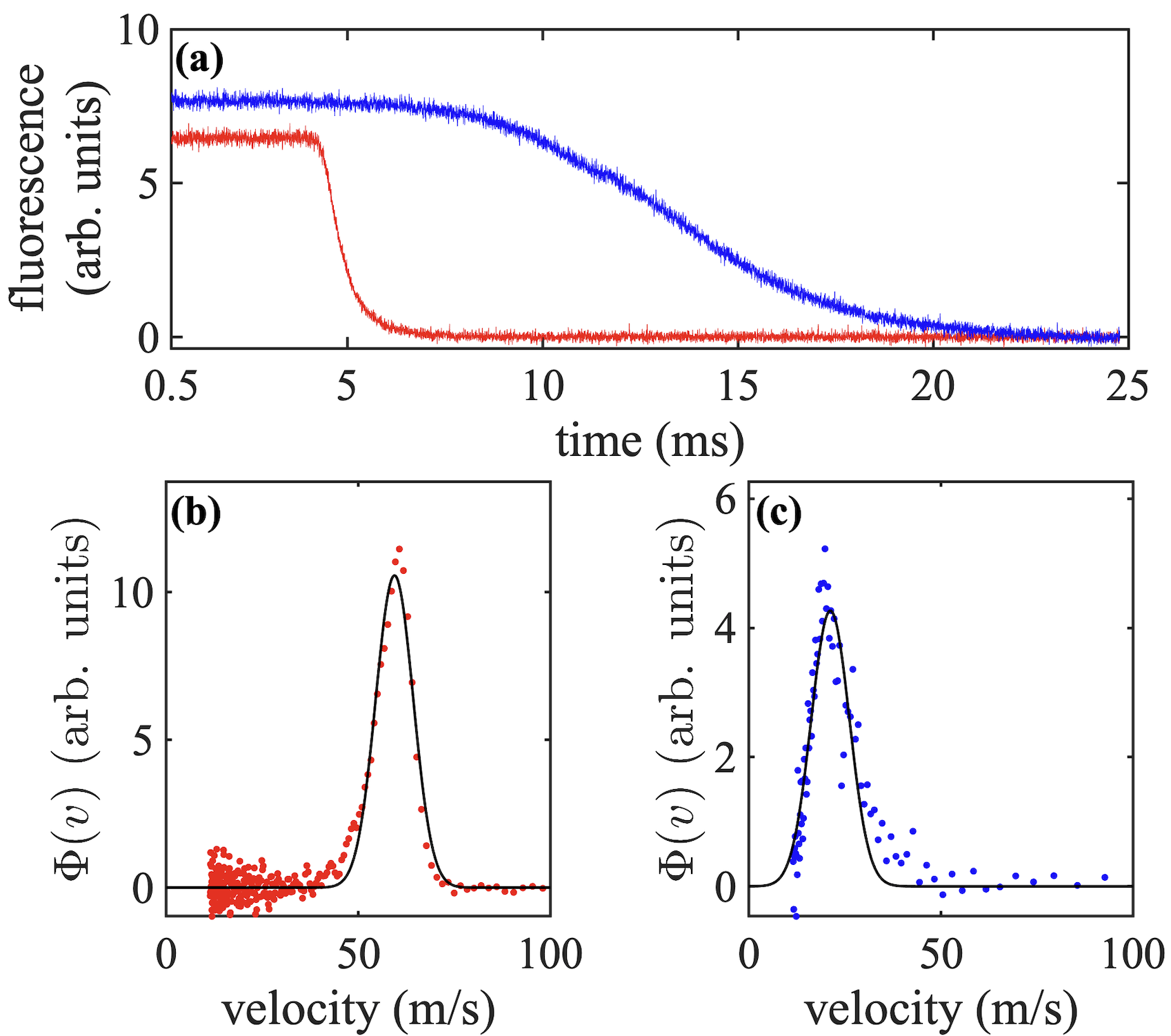}
    \caption{(Color Online) (a) Fluorescence signals of Li (red) and Cs (blue) as function of time measured after the 2D MOT is shut off abruptly. The velocity distributions of the beam flux $\Phi(v)$ for Li~(b) and Cs~(c) are derived from taking a discrete derivative of the fluorescence signals as shown in (a). Black lines are Gaussian fits to the data. The most-probable velocities of Li and Cs are 57 and 21~m/s, respectively.}
    \label{fig:TOF_dist}
\end{figure}

In a TOF realization, we record the decay of the fluorescence until all atoms pass the probing region. We average over 500 realizations in a typical measurement to suppress noise. Figure~\ref{fig:TOF_dist}(a) shows a typical Li(Cs) measurement with pushing beam intensity 14.1(2.0)~mW/cm$^{2}$ and detuning 8 $\Gamma_{\mathrm{Li}}$(1 $\Gamma_{\mathrm{Cs}}$). The velocity distribution of the beam flux can be derived from the discrete derivative of the fluorescence signal \cite{PhysRevA.58.3891}, as shown in Figs.~\ref{fig:TOF_dist}(b) and 3(c). We fit the velocity distributions to a Gaussian to obtain the most-probable velocities.

\begin{figure}[h]
    \centering
    \includegraphics[width=0.4\textwidth]{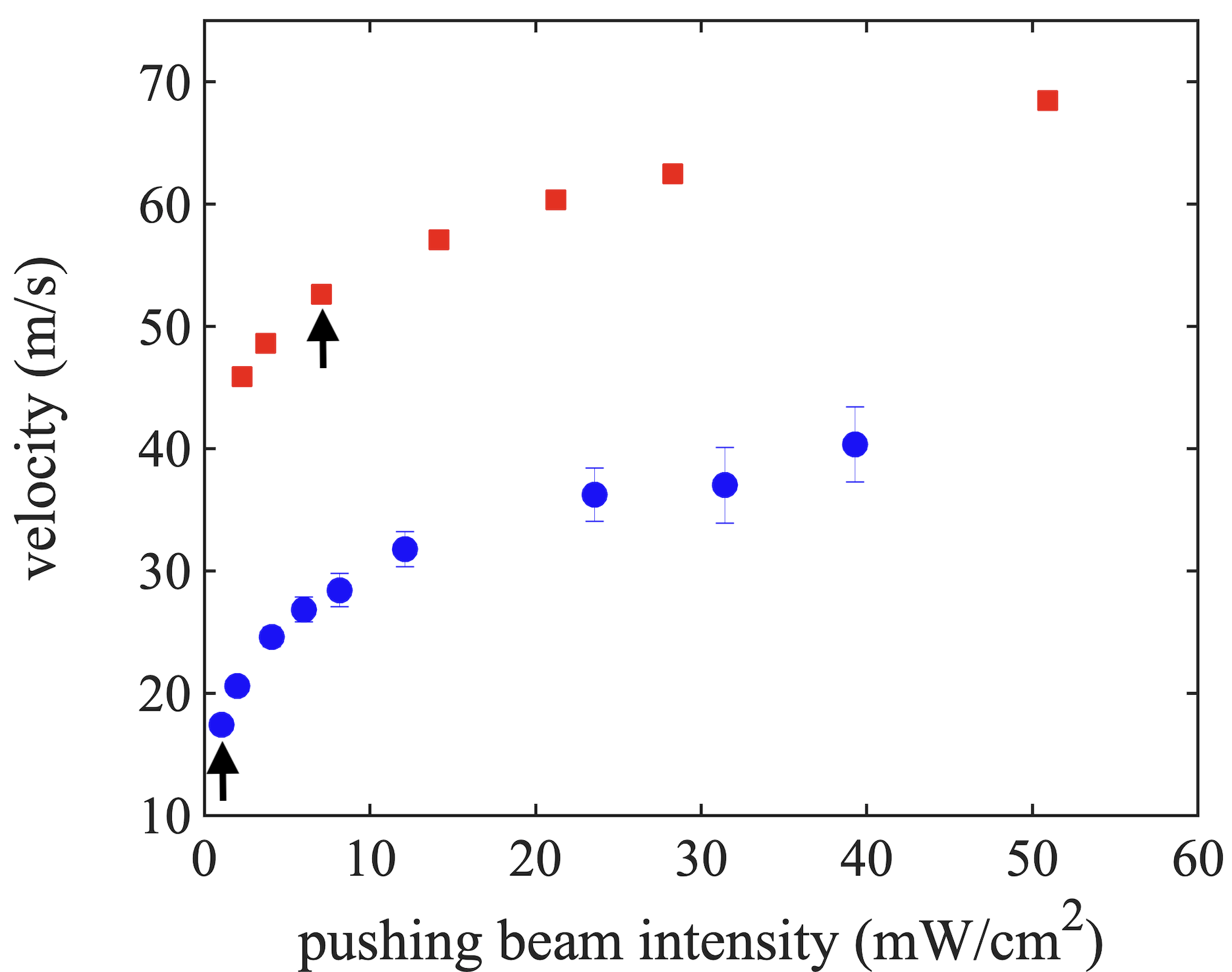}
    \caption{(Color Online) The most-probable velocities of Li (red squares) and Cs (blue circles) as a function of pushing beam intensity. The Li(Cs) data are taken with a pushing beam detuning of~8 $\Gamma_{\mathrm{Li}}$(1 $\Gamma_{\mathrm{Cs}}$). The error bars are smaller than the marker size in the Li data. The two black arrows indicate the Li and Cs velocities that give the maximum Li and Cs loading rates, respectively.}
    \label{fig:Most-probable velocity} 
\end{figure}

The most-probable velocity as a function of pushing beam intensity is plotted in Fig.~\ref{fig:Most-probable velocity}. The most-probable velocity increases with the intensity of the pushing beam. Experimentally, we vary the intensities of the pushing beams to control the most-probable velocities of the slow beam.

\subsection{Pushing beam optimization}

\begin{figure}[h]
    \centering
    \includegraphics[width=0.4\textwidth]{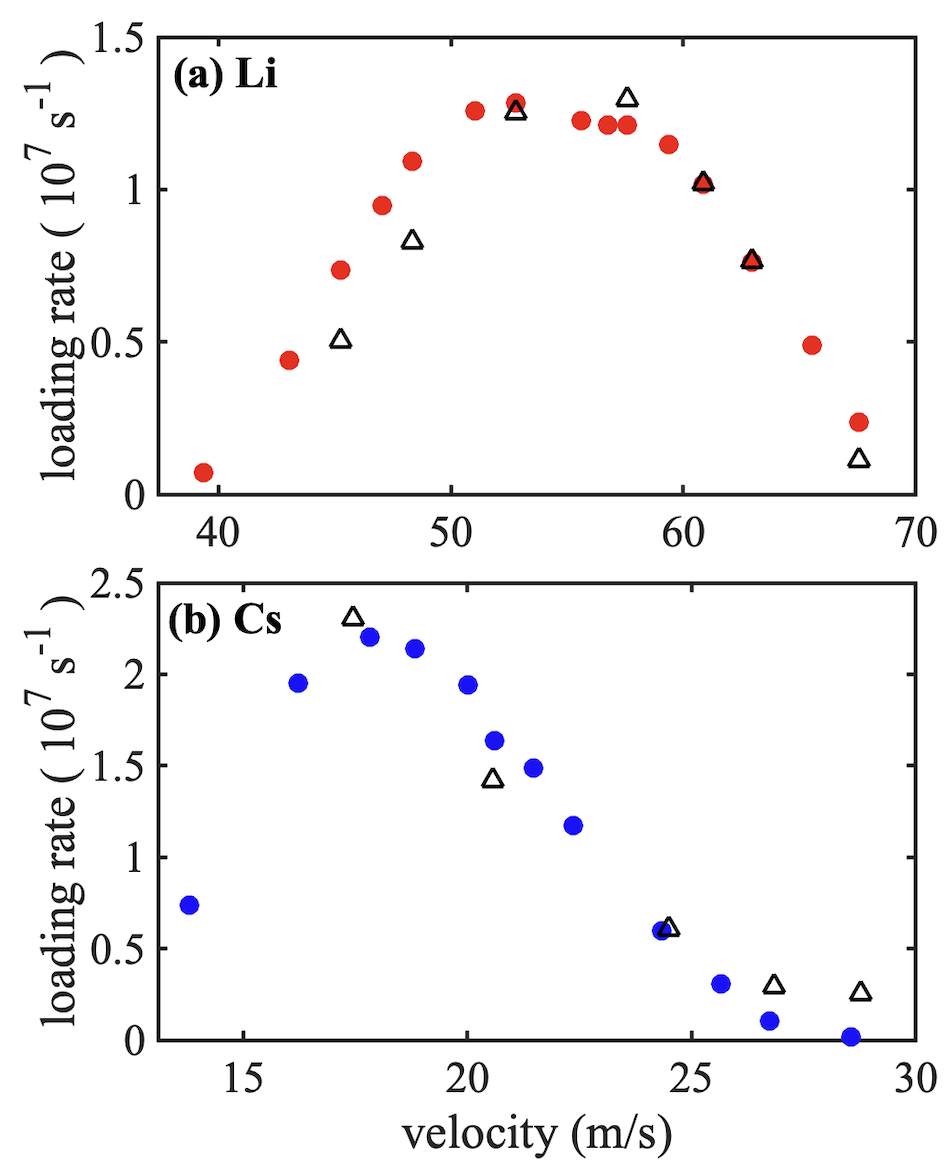}
    \caption{(Color Online) 3D MOT loading rates of Li (a) and Cs (b) as a function of the most-probable velocity. The measurements are taken with a fixed 2D MOT field gradient of 31~G/cm. The open triangles are the fits to the data using the velocity distributions of the slow beam obtained from the derivative of the fluorescence decay signals. To find the best fits, we vary the capture velocity of the 3D MOT and the constant that converts fluorescence to atom number.  }
    \label{fig:3D MOT loading rate}
\end{figure}

The longitudinal velocity distribution of the slow beam plays an important role in the 3D MOT loading. Figure~\ref{fig:3D MOT loading rate} shows the measured 3D MOT loading rate as a function of the most-probable velocity. The Li and Cs loading rates both increase with the pushing beam intensities initially and reach their peak values at the velocities around 52~m/s and 18~m/s, respectively. As the intensities increase further, the loading rates start to decrease. This observation can be interpreted as the velocity of atoms in the atomic beam starts to exceed the capture velocity of the 3D MOT.

To understand the behavior of 3D MOT loading, we extract the velocity distributions of the beam fluxes from the fluorescence signals measured with different pushing beam intensities, as shown in Fig.~\ref{fig:TOF_dist}. Then, each of the distributions is summed up to a capture velocity to obtain the flux captured in the 3D MOT. We vary the value of capture velocity and the value of an overall number scaling constant to obtain the best fits to the measured loading rates shown in Fig.~\ref{fig:3D MOT loading rate}. Our best fits yield a capture velocity of  62~m/s for Li and 17~m/s for Cs. Based on a simple one-dimensional model, the 3D MOT capture velocities of Li and Cs are found to be 51 and 18 m/s, respectively. The capture velocities derived from two different methods are in good agreement.

Atoms in the atomic beam must have $v_z\gtrsim 11$~m/s to avoid dropping below the 3D MOT trapping region because of gravity. When the pushing beams are absent, the Li fluorescence from the 3D MOT disappears, but a small amount of Cs fluorescence is still present. This observation can be explained by the very different longitudinal velocity distributions of Li and Cs atoms in the 2D MOT without pushing beams. The Li atoms from the bottom of the oven must have a $v_z$ to $v_y$ ratio lower  than~0.1 to avoid sticking to the wall of the oven tube. The 2D MOT capture velocity is estimated to be 50~m/s, leading to $v_z$=5~m/s; thus, the largest possible longitudinal velocity of Li in the 2D MOT is smaller than 11~m/s. In contrast to Li, Cs is a volatile species and has a vapor pressure of $\sim$$10^{-6}$~mbar around room temperature. In this case, the Cs atoms entering the 2D MOT are not very directional, so a small amount of Cs atoms with a large enough $v_z$ can arrive at the 3D MOT without a pushing beam. 

Because Cs is a highly volatile species, a Cs vapor leakage to the quartz cell can degrade the vacuum inside. In experiments, we use the decay of the Li 3D MOT to assess the degradation. In the absence of the Cs 3D MOT, we measure decay time constants of the Li 3D MOT in two different conditions: (1) with the Cs 2D MOT and the Cs pushing beam and (2) without the Cs 2D MOT and the Cs pushing beam.  In both conditions, the trap decay measurements yield similar time constants of $\sim$150~s; thus, the degradation of vacuum because of the Cs leakage should be well below the level of $10^{-11}$~torr.


\subsection{2D MOT optimization}
The 2D MOT performance is optimized by several parameters such as the magnetic field gradient, the power and detuning of the Cs trapping beams, and the detuning of the Li trapping and repumping beams. We use the fluorescence from the atoms passing the resonant probe beam in the second chamber to quantify the performance of the 2D MOT.

\begin{figure}[h]
\centering
\includegraphics[width=0.5\textwidth]{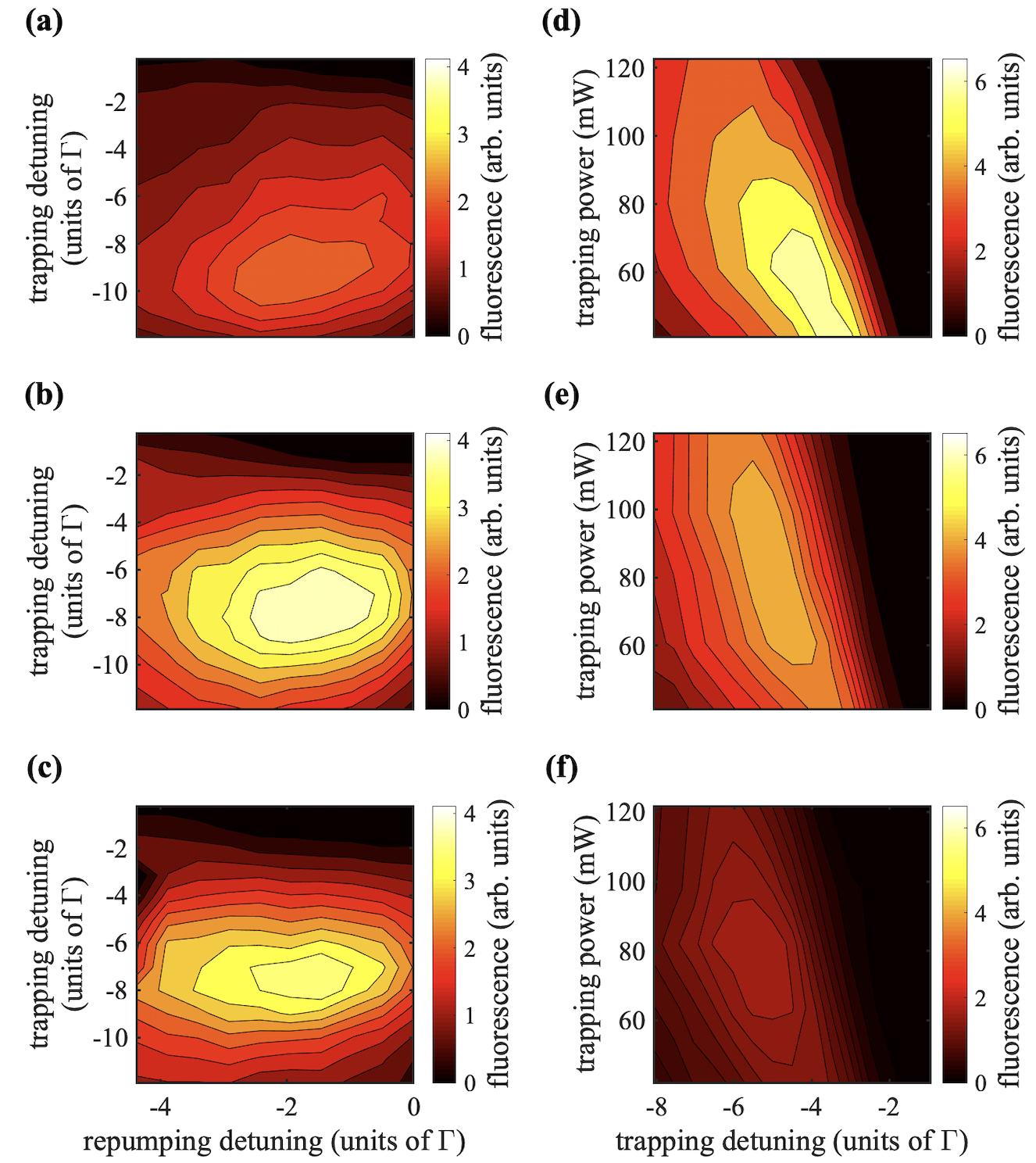}
\caption{(Color Online) 2D MOT optimization for Li (a-c) and Cs (d-f): for Li, the fluorescence strength as a function of the detunings of the Li trapping and repumping light is measured at three different magnetic field gradients: (a) 17, (b) 31, and (c) 40~G/cm. For Cs, the fluorescence strength as a function of the detuning of the Cs trapping light and the power in each Cs trapping beam is investigated at a field gradient of (d) 17, (e) 31, and (f) 40~G/cm .}
\label{fig:2D MOT scan} 
\end{figure}

Among the parameters, the magnetic field gradient is the only parameter shared by both species; other parameters can be adjusted independently. We choose three magnetic field gradient values, 17, 31, and 40~G/cm, for the optimization. With each field gradient value, we investigate the Li fluorescence strength as a function of the detunings of the Li trapping and repumping beams as well as the Cs fluorescence strength as a function of the Cs trapping beam power and detuning. The results are shown in Fig.~\ref{fig:2D MOT scan}. 

\begin{figure}[h]
    \centering
    \includegraphics[width=0.4\textwidth]{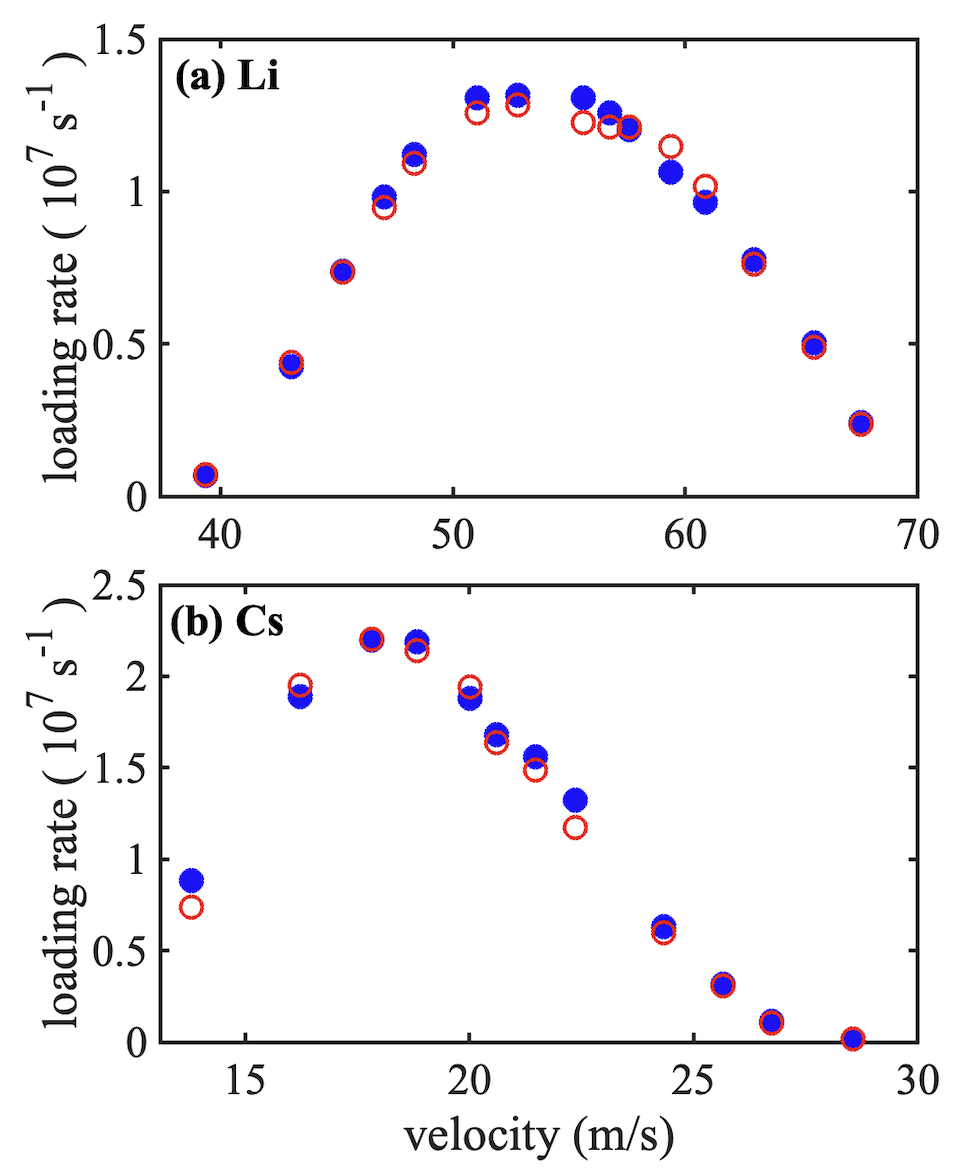}
    \caption{(Color Online) 3D MOT loading rate as a function of the most-probable velocity. In (a) and (b), we compare the Li and Cs loading rates measured with (solid circles) and without (open circles) the second-species atoms, respectively. The data represented by the open circles are identical to those shown in Fig.~\ref{fig:3D MOT loading rate}.}
    \label{fig:Interaction between different species at 2D MOT}
\end{figure}

Comparing the results from the three field gradients, we achieve the absolute maximum Li 2D MOT output with a magnetic gradient of 31~G/cm, while the absolute maximum Cs output is obtained with a magnetic gradient of 17~G/cm. The best value of the Li output found at a magnetic field gradient of 17~G/cm is about 57\% of the best output at 31~G/cm. For Cs, the best output found at a magnetic field gradient of 31~G/cm is about 68\% of the best value at 17~G/cm. With the magnetic field gradient of the 2D MOT fixed at 31~G/cm, we achieve the best 3D MOT loading rates of $1.3\times10^7$~s$^{-1}$ (Li) and 2.2$\times10^7$~s$^{-1}$ (Cs). These best loading rates are obtained with the following 2D MOT beam parameters: the Li trapping detuning $\delta_{\mathrm{t}}^{\mathrm{Li}}$=-7~$\Gamma_{\mathrm{Li}}$, the Li repumping detuning $\delta_{\mathrm{r}}^{\mathrm{Li}}$=-1.5~$\Gamma_{\mathrm{Li}}$, the Cs trapping detuning $\delta_{\mathrm{t}}^{\mathrm{Cs}}$=-4.5~$\Gamma_{\mathrm{Cs}}$, and the Cs repumping detuning $\delta_{\mathrm{r}}^{\mathrm{Cs}}$=0. The trapping and repumping beam intensity of Li(Cs) are 10.5(18.7) and 5.8(1.1)~mW/cm$^{2}$; the data shown in Figs~\ref{fig:TOF_dist},~\ref{fig:Most-probable velocity},~\ref{fig:3D MOT loading rate}, and ~\ref{fig:Interaction between different species at 2D MOT} are taken with these intensities.

At the same Li oven temperature, our best Li loading rate is about a factor 20 smaller than the one reported in Ref.~\cite{PhysRevA.80.013409}. Except for the compromised field gradients, the distance between the 2D and the 3D MOT in our system is two times longer than the one in Ref.~\cite{PhysRevA.80.013409}, giving rise to a factor of four reduction in the flux density at the location of the 3D MOT. Similar factors also diminish the Cs loading rates. Compared with the existing Cs source reported in Ref.~\cite{PhysRevA.60.R4241}, our best Cs loading rate is a factor of six smaller.

We also check if the two overlapped beams can interfere. The 3D MOT loading rates as a function of the most-probable velocity are measured with and without the presence of the other species. As shown in Fig.~\ref{fig:Interaction between different species at 2D MOT}, we do not observe any differences outside the measurement uncertainties. This indicates the possibility of including more species into the current 2D MOT setup without major modifications.

\section{Conclusion}

We have realized a two-species 2D MOT that produces simultaneous and overlapped slow beams of lithium and cesium. For a Li-only and Cs-only operation, we obtain a maximum 3D MOT loading rate of $1.3\times10^7$ s$^{-1}$ for Li and $2.2\times10^7$ s$^{-1}$ for Cs, resulting in a saturation number of $1.0\times10^{9}$ Li atoms and $1.4\times10^{8}$ Cs atoms in the 3D MOT, with the Li oven temperature at 370$^{\circ}$C and the Cs oven temperature at 20$^{\circ}$C.

The optimization of the two-species atomic beam is straightforward, because the parameters associated with individual species are mostly decoupled in the 2D MOT. The only exception is the quadrupole magnetic gradient of the 2D MOT. Despite being coupled by the field gradient, the beam fluxes do not have a strong dependence near the optimum value. The other parameters such as laser detunings and intensities can be independently tuned to optimize the 3D MOT loading rate. 
The pushing beams' parameters are found to be the most crucial, since they determine the longitudinal velocity distribution and the pointing of the atomic beam. 

Furthermore, we do not observe any interference between the Li and Cs atoms in the beam. This can be explained by  the fact that the pushing beams constantly push atoms out of the 2D MOT and keep the densities of the trapped atoms low. With straightforward optimization and negligible interference between different species, it is likely that a multi-species 2D MOT adopting a similar design can also generate satisfactory fluxes for future applications.\\

\begin{acknowledgements}
The authors thank G. Lamporesi for many fruitful discussions. This work was supported by the Ministry of Science and Technology of Taiwan (Grants No.105-2112-M-007-037-MY3, No.108-2112-M-007-013, and No.109-2112-M-007-021). We further acknowledge funding from the Ministry of Education of Taiwan and the Foundation for the Advancement of Outstanding Scholarship.
\end{acknowledgements}

\end{document}